\documentclass[conference]{IEEEtran}
\IEEEoverridecommandlockouts

\usepackage{cite}
\usepackage{amsmath,amssymb,amsfonts}
\usepackage{graphicx,color}
\graphicspath{{fig/}}  
\DeclareGraphicsExtensions{.pdf,.png,.jpg,.jpeg}
\usepackage{subcaption}
\usepackage[export]{adjustbox}
\usepackage{textcomp}
\usepackage[hyphens]{url}          
\usepackage[table]{xcolor}
\usepackage{booktabs}
\usepackage{paralist}
\usepackage{enumitem}
\usepackage{multirow}
\usepackage{hyperref}
\usepackage{algorithmic}
\usepackage[linesnumbered,ruled,vlined]{algorithm2e}
\usepackage{float}                 
\usepackage{amssymb}               
\usepackage[section]{placeins}     

\newcommand{\xmark}{\text{\sffamily X}} 
\hypersetup{colorlinks=true,linkcolor=red,citecolor=blue,urlcolor=magenta}
 \def\eg{{\em e.g.}}
\def\BibTeX{{\rm B\kern-.05em{\sc i\kern-.025em b}\kern-.08em
    T\kern-.1667em\lower.7ex\hbox{E}\kern-.125emX}}

\newcommand{\smallurl}[1]{\footnotesize\url{#1}}
\definecolor{baselinecolor}{gray}{.9}

\usepackage{pifont}
\newcommand{\halfcheckmark}{\ding{51}\hspace{-1.85mm}\ding{55}}
\newcommand{\rulesep}{\unskip\ \vrule\ }

\begin{document}

\title{SynthGuard: An Open Platform for Detecting AI-Generated Multimedia with Multimodal LLMs}

\author{\IEEEauthorblockN{Shail Desai\textsuperscript{1}, Aditya Pawar\textsuperscript{1},
Li Lin\textsuperscript{1},
Xin Wang\textsuperscript{2},  Shu Hu\textsuperscript{1}$^*$\thanks{$^*$Corresponding Author}}
\IEEEauthorblockA{\textsuperscript{1}Purdue University, West Lafayette, IN, USA {\small \{shdesai, aspawar, lin1785, hu968\}@purdue.edu}\\
\textsuperscript{2}University at Albany, State University of New York, Albany, NY, USA {\small xwang56@albany.edu}}}

\maketitle

\thispagestyle{plain}
\pagestyle{plain}

\begin{abstract}
Artificial Intelligence (AI) has made it possible for anyone to create images, audio, and video with unprecedented ease, enriching education, communication, and creative expression. At the same time, the rapid rise of AI-generated media has introduced serious risks, including misinformation, identity misuse, and the erosion of public trust as synthetic content becomes increasingly indistinguishable from real media. Although deepfake detection has advanced, many existing tools remain closed-source, limited in modality, or lacking transparency and educational value, making it difficult for users to understand how detection decisions are made. To address these gaps, we introduce \textit{SynthGuard}, an open, user-friendly platform for detecting and analyzing AI-generated multimedia using both traditional detectors and multimodal large language models (MLLMs). SynthGuard provides explainable inference, unified image and audio support, and an interactive interface designed to make forensic analysis accessible to researchers, educators, and the public. The SynthGuard platform is available at: \url{https://in-engr-nova.it.purdue.edu/}.  
\end{abstract}



\section{Introduction}

AI has transformed how people create and experience media~\cite{lin2024detecting}. Modern AI tools (\eg, Sora 2 \cite{sora2}, Nano Banana \cite{nano}) can now generate images, videos, and audio content that once required advanced skills or costly equipment~\cite{lin2024detecting,chui2022generative}. These technologies are being used to enrich daily life—enabling educators to design engaging lessons~\cite{malinka2023educational}, artists to visualize concepts, and small businesses to produce professional-quality marketing materials~\cite{chui2022generative}. By lowering creative barriers, AI-generated media has made content creation faster, more accessible, and more inclusive, fostering innovation across education, entertainment, and communication~\cite{lin2024detecting}.

However, the same accessibility that empowers creativity also introduces serious challenges. AI-generated media, often referred to as \textit{deepfakes}, can blur the line between reality and fabrication~\cite{lin2024detecting,wang2024spotting}. Such content can be misused to spread misinformation, impersonate individuals, or manipulate public perception~\cite{chen2024self,chen2023harnessing,zheng2024few,lin2025fit}. As synthetic media becomes increasingly realistic and widespread, the ability to identify, verify, and understand AI-generated content is essential for preserving public trust and digital integrity~\cite{lin2024detecting}.

Despite recent progress in deepfake detection~\cite{wang2024spotting,ren2024improving,krubha2025robust,lin2024robust1}, existing tools are often closed-source, difficult to use, or limited in modality~\cite{deepware2022,ju2024deepfakeometerv20openplatform}. Many lack explainability, accessibility, or educational value—preventing users from understanding how and why a piece of media is flagged as synthetic~\cite{guo2022open,guo2022eyes}. There remains a critical need for a free, user-friendly, and transparent detection platform that integrates explainable AI capabilities through multimodal large language models (MLLMs), bridging the gap between technical research and real-world understanding~\cite{lin2024detecting,hou2025rethinking,lin2025ai,lin2024preserving,ju2024improving}.

\begin{figure*}[!t]
  \centering
  \includegraphics[width=1\linewidth]{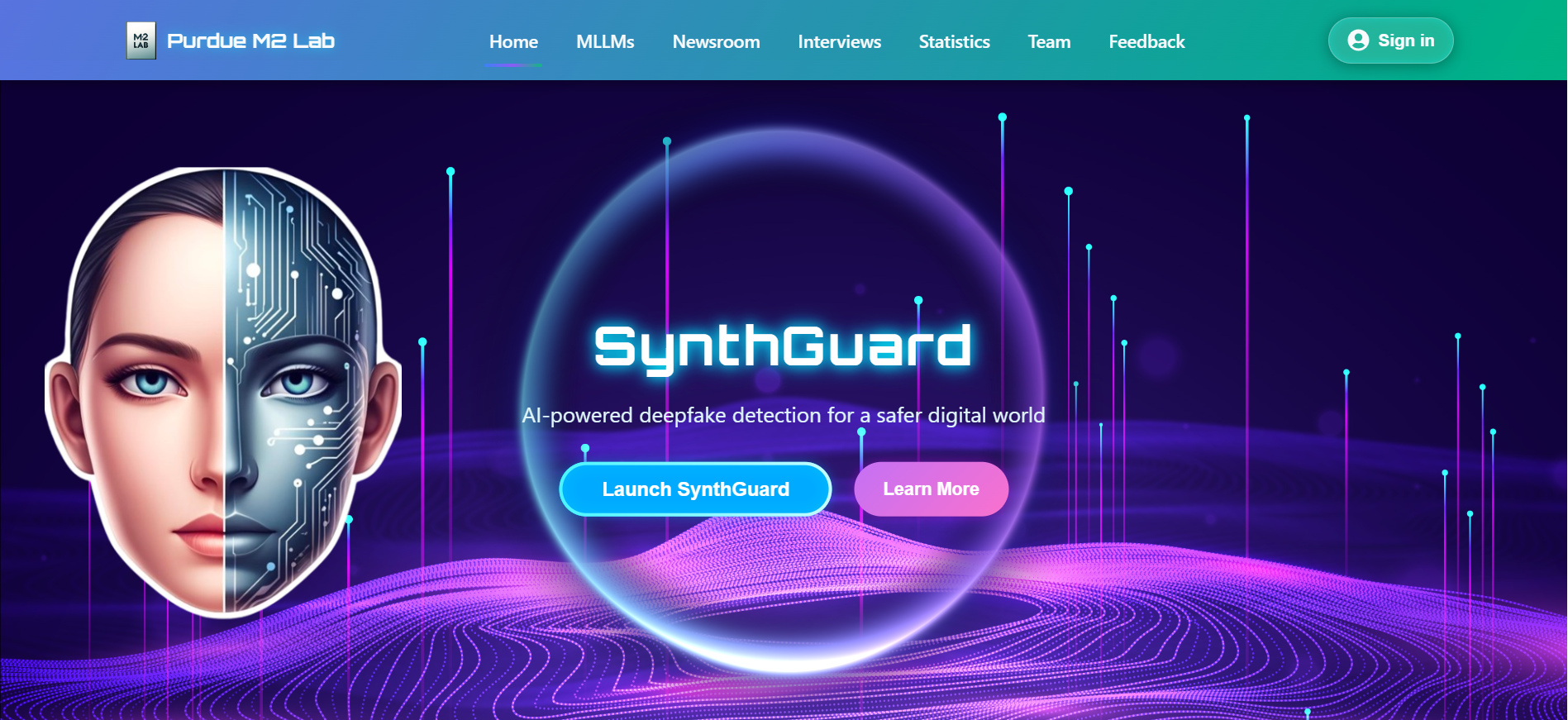}
  \caption{SynthGuard UI overview.}
  \label{fig:ui}
\end{figure*}

In this work, we present \textbf{SynthGuard} (see Fig.~\ref{fig:ui}), an open and interactive platform for detecting and analyzing AI-generated media that makes forensic analysis explainable and accessible to both researchers and the general public. The system couples an advanced, UX-oriented web interface with a modularized Python backend, yielding a data-driven web application for coordinated detection, logging, and MLLM explanation.
Our contributions are summarized as follows:
\begin{enumerate}[leftmargin=*,noitemsep]
    \item A new publicly available platform is developed for the detection of AI-generated multimedia content, encompassing both images and audio.
    \item Our platform is the first open MLLM-based explainable platform for detecting AI-generated multimedia, capable of providing reasoning for its detection results.
\end{enumerate}

\section{Related Work}

\begin{table*}[t]
\centering
\caption{Summary of existing platforms for AI-generated media detection.
A checkmark (\checkmark) denotes full/free support; a \halfcheckmark denotes a limited or trial-based free tier; and an \xmark\ denotes no free support. ``MLLM-Aware'' means the system supports multimodal large language models for assisted forensic reasoning.}
\label{tab:platforms}
\renewcommand{\arraystretch}{1.2}
\begin{tabular}{c|c|cc|c|c}
\hline
\multirow{2}{*}{\textbf{Platform}} & \multirow{2}{*}{\textbf{Free}} &
\multicolumn{2}{c|}{\textbf{MLLM-Agnostic Detector}} &
\multirow{2}{*}{\textbf{MLLM-Aware Detector}} &
\multirow{2}{*}{\textbf{Website}} \\
\cline{3-4}
 &  & \textbf{Image} & \textbf{Audio} &  &  \\ \hline\hline

Deepware & \checkmark & \checkmark & \xmark & \xmark & \smallurl{https://scanner.deepware.ai/} \\ \hline
DuckDuckGoose & \xmark & \checkmark & \xmark & \xmark & \smallurl{https://duckduckgoose.ai/} \\ \hline
Sensity AI & \xmark & \checkmark & \xmark & \xmark & \smallurl{https://sensity.ai/deepfake-detection/} \\ \hline
Resemble.AI & \halfcheckmark & \checkmark & \checkmark & \xmark & \smallurl{https://resemble.ai/free-deepfake-detector/} \\ \hline
TrueMedia & \checkmark & \checkmark & \checkmark & \xmark & \smallurl{https://truemedia.org/} \\ \hline
Pindrop Pulse & \xmark & \xmark & \checkmark & \xmark & \smallurl{https://www.pindrop.com/deepfake/} \\ \hline
DeepFake-o-meter v2.0 & \checkmark & \checkmark & \checkmark & \xmark & \smallurl{https://zinc.cse.buffalo.edu/ubmdfl/deep-o-meter} \\ \hline
Incode & \xmark & \checkmark & \xmark & \xmark & \smallurl{https://incode.com/} \\ \hline
Is It AI & \halfcheckmark & \checkmark & \xmark & \xmark & \smallurl{https://isitai.com/} \\ \hline
Winston AI & \halfcheckmark & \checkmark & \xmark & \xmark & \smallurl{https://gowinston.ai/} \\ \hline
BrandWell & \xmark & \checkmark & \xmark & \xmark & \smallurl{https://brandwell.ai/ai-image-detector/} \\ \hline
Illuminarty & \halfcheckmark & \checkmark & \xmark & \xmark & \smallurl{https://illuminarty.ai/en/} \\ \hline
Reality Defender & \halfcheckmark & \checkmark & \xmark & \xmark & \smallurl{https://www.realitydefender.com/} \\ \hline
Hive Moderation & \halfcheckmark & \checkmark & \xmark & \xmark & \smallurl{https://hivemoderation.com/} \\ \hline
Deepfake Detector & \xmark & \checkmark & \xmark & \xmark & \smallurl{https://deepfakedetector.ai/} \\ \hline\hline
\textbf{SynthGuard (Ours)} & \checkmark & \checkmark & \checkmark & \checkmark & \smallurl{https://in-engr-nova.it.purdue.edu/} \\ \hline
\end{tabular}
\end{table*}

\subsection{AI-Generated Media}
Artificial Intelligence has become a creative partner in people’s everyday lives. With modern tools, anyone can generate images, videos, and audio clips that once required professional equipment or specialized skills \cite{lin2024detecting}. For example, a designer can instantly visualize ideas, an educator can create short animated lessons \cite{malinka2023educational}, and a small business owner can record a natural-sounding voiceover without hiring a studio \cite{chui2022generative}. Students now use AI to illustrate reports, musicians experiment with AI-assisted composition, and filmmakers prototype scenes before production. Across creative fields, AI has made it possible to transform imagination into media with just a few words or sentences.

The impact of this technology extends beyond convenience. AI-generated media is enriching communication, accessibility, and artistic expression. It allows people to tell stories more vividly, translate content into multiple languages, and create personalized material for audiences around the world. For individuals with disabilities, text-to-speech and image-description tools provide new ways to interact with digital information. For others, AI art and video generation lower barriers to creativity, making complex ideas easier to share and understand. By making creation faster, cheaper, and more inclusive, AI-generated media is changing how people learn, express themselves, and connect with others.

Yet, the same ease of generation that empowers creativity also introduces challenges. The line between real and synthetic content is becoming increasingly thin, raising questions about authenticity, privacy, and consent. As AI-produced material spreads across social media \cite{chen2024self,chen2023harnessing} and communication platforms \cite{zheng2024few}, the need to recognize and verify its origins becomes essential to maintaining trust in digital information \cite{lin2025fit}.

\subsection{Detecting AI-Generated Media}
Detecting AI-generated media means determining whether a photo, video, or audio recording was created or altered by artificial intelligence \cite{wang2024spotting}. This task has grown in importance as AI tools become more powerful and widely available. People can now produce lifelike voices \cite{ren2024improving}, realistic faces, or fabricated news clips that appear genuine. While these capabilities enhance creativity and entertainment, they can also be misused for misinformation, identity theft, or political manipulation. Understanding when content is artificially generated is therefore critical for protecting individuals and preserving trust in public communication.

The goal of detection is not to restrict the use of AI but to support transparency \cite{guo2022open, guo2022eyes,wang2023gan,hu2021exposing}, robustness \cite{krubha2025robust, lin2024robust1, zhang2024x, chen2024masked,fan2024synthesizing,yang2025crossdf,fan2023attacking,pu2022learning,guo2022robust}, and fairness \cite{hou2025rethinking, wu2025preserving, lin2025ai,lin2024preserving,ju2024improving}. By identifying AI-created material, researchers, journalists, and everyday users can better understand the reliability of what they see and hear online. Effective detection helps prevent deception, ensures fair representation, and allows audiences to appreciate the benefits of AI without fear of manipulation. In this way, the study of AI-generated media and its detection forms a necessary balance—celebrating innovation while safeguarding truth.

\subsection{Existing Detection Platforms}
Table~\ref{tab:platforms} shows that most existing platforms are either fully paid or restricted to narrow trial-based tiers, and coverage is highly fragmented: several systems focus solely on images, while only a few (Resemble.AI, TrueMedia, DeepFake-o-meter v2.0) offer both image and audio support. None of the surveyed platforms provide native MLLM-aware analysis for interactive, explainable forensics. {SynthGuard} is designed to close this gap by offering full/free access in our research deployment, unified image--audio deepfake detection, and integrated MLLM-based reasoning.



\section{SynthGuard Frontend}
\begin{figure*}[t]
\captionsetup[subfigure]{justification=centering}
\centering
        \begin{subfigure}[b]{0.4\textwidth}
                \includegraphics[width=\linewidth]{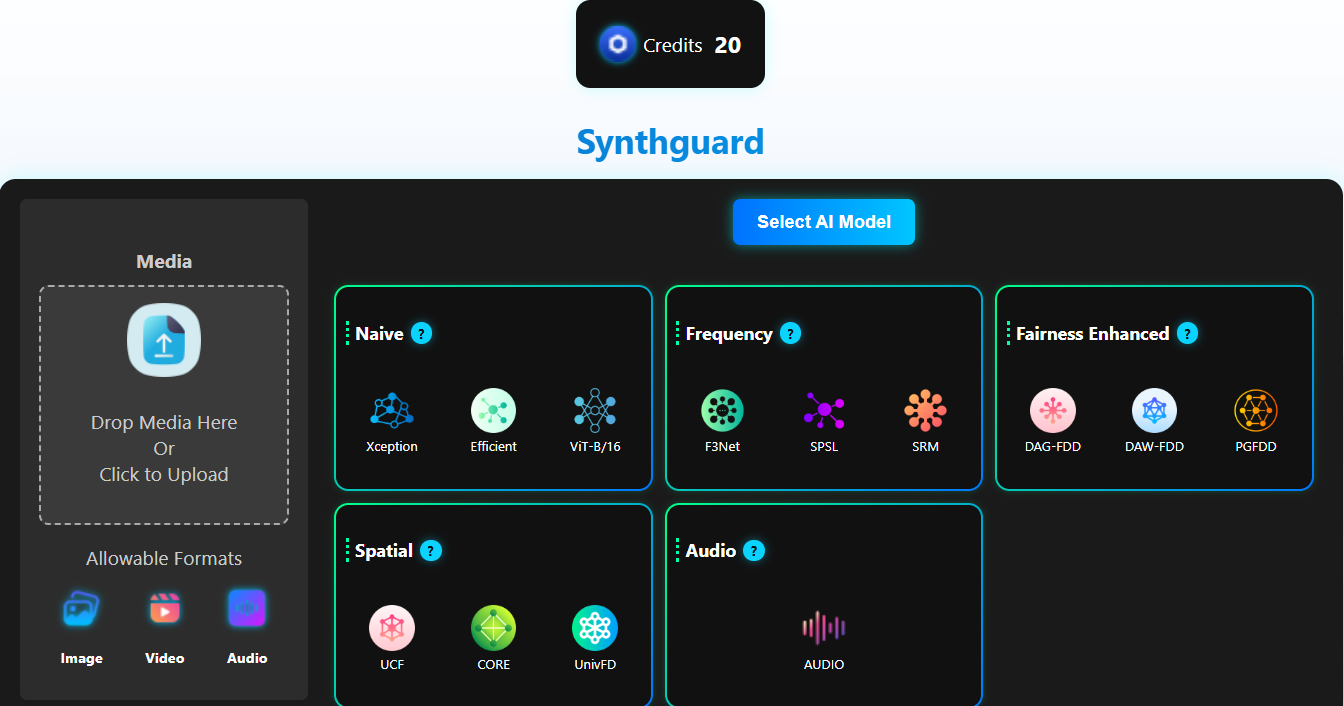}
                \caption{}
                \label{fig:LogisticRegression_data4_2}
        \end{subfigure}%
        \rulesep
        \begin{subfigure}[b]{0.58\textwidth}
                \includegraphics[width=\linewidth]{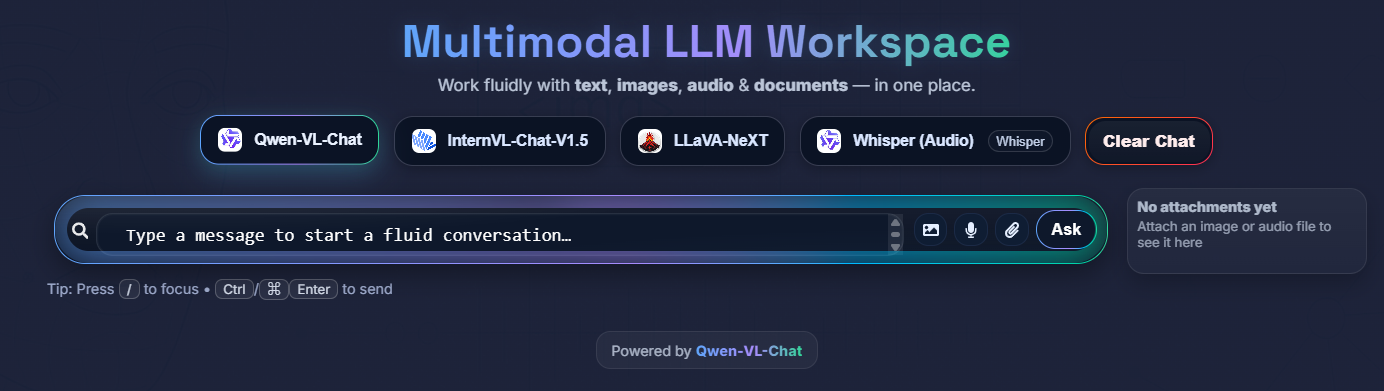}
                \caption{}
                \label{fig:LogisticRegression_data4_3}
        \end{subfigure}%
        
        \begin{subfigure}[b]{0.485\textwidth}
                \includegraphics[width=\linewidth]{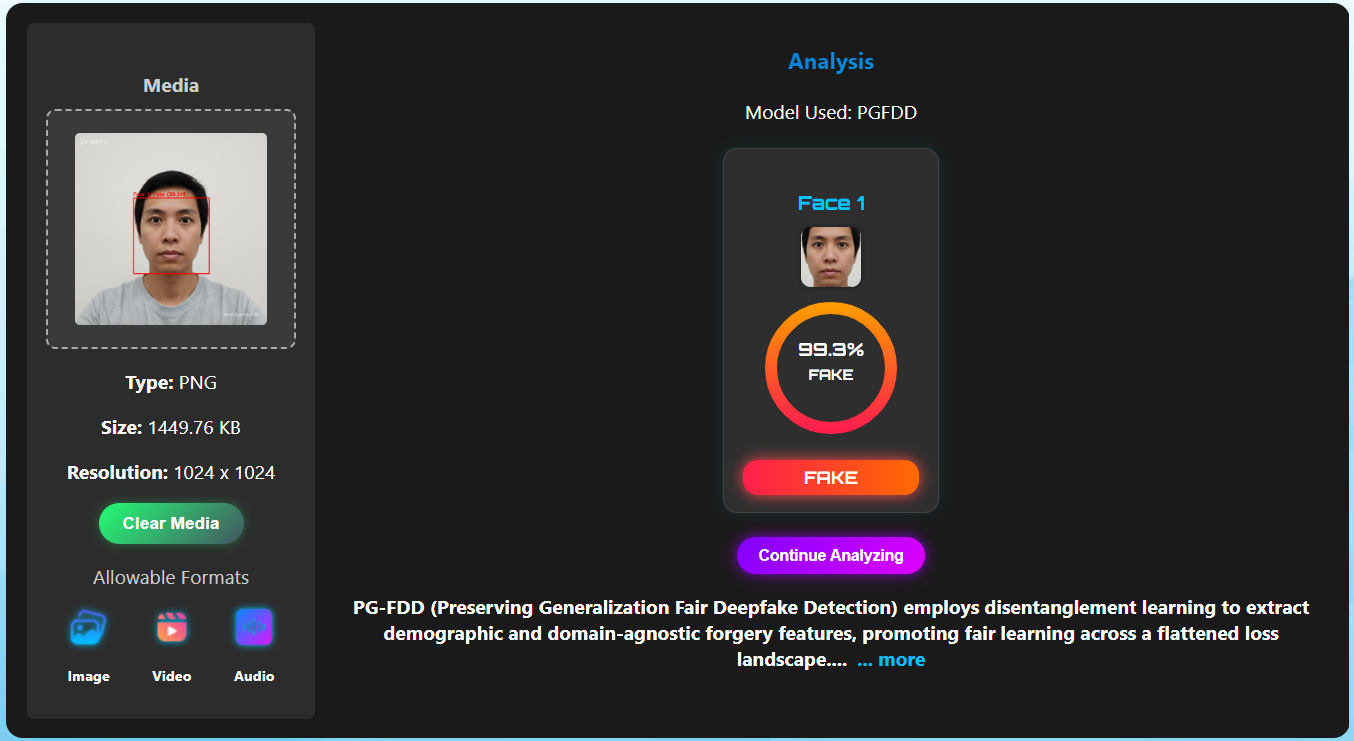}
                \caption{}
                \label{fig:LogisticRegression_data4_4}
        \end{subfigure}%
        \rulesep
        \begin{subfigure}[b]{0.49\textwidth}
                \includegraphics[width=\linewidth]{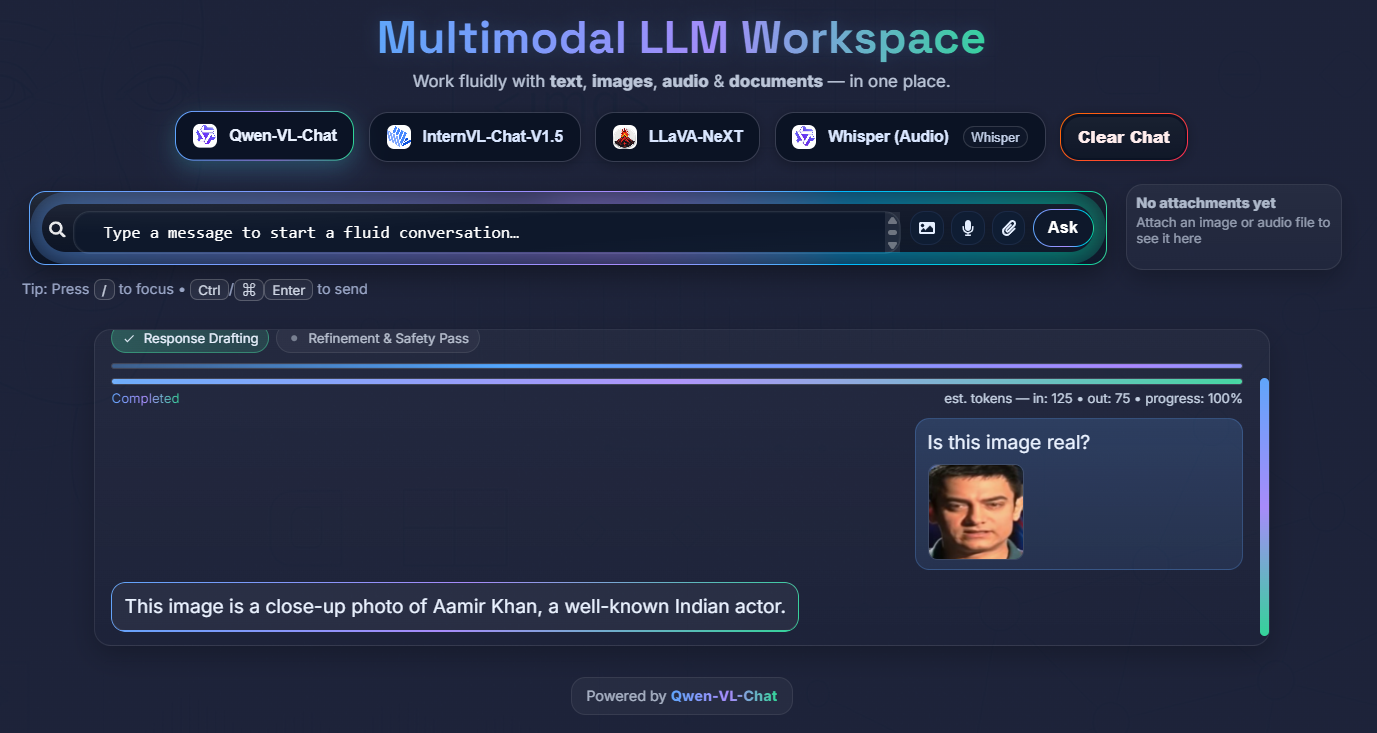}
                \caption{}
                \label{fig:LogisticRegression_data4_5}
        \end{subfigure}%
        
        \begin{subfigure}[b]{0.49\textwidth}              \includegraphics[width=\linewidth]{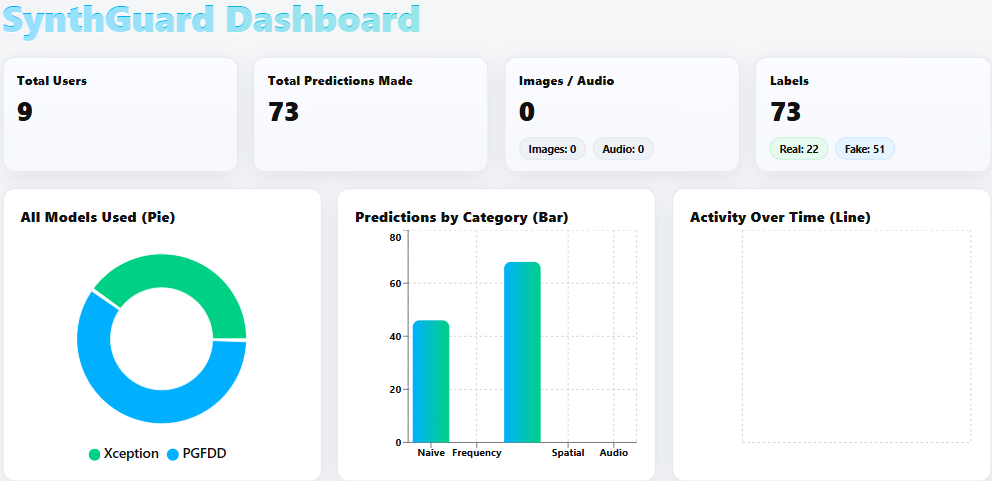}
                \caption{} 
                \label{fig:LogisticRegression_data4_10}
        \end{subfigure}%
        \rulesep
        \begin{subfigure}[b]{0.49\textwidth}
                \includegraphics[width=\linewidth]{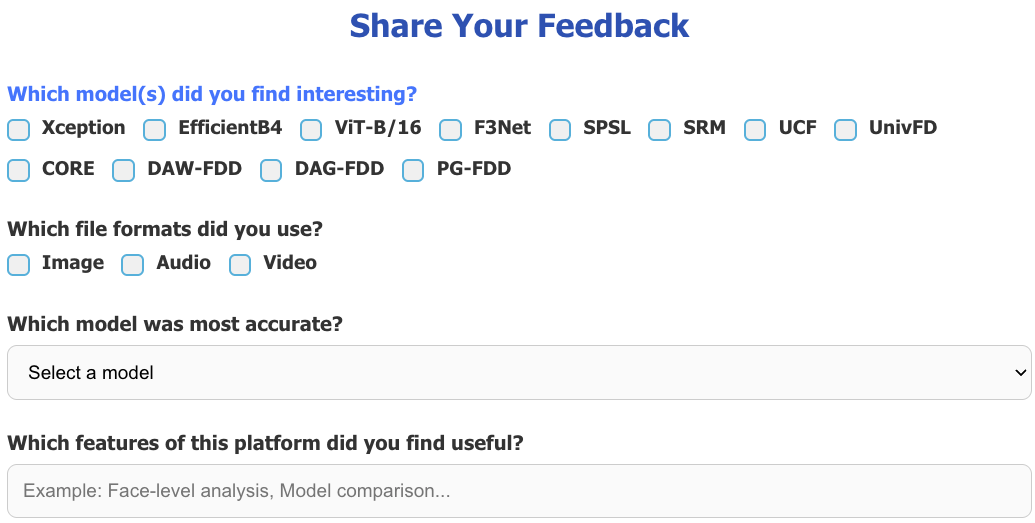}
                \caption{}
                \label{fig:LogisticRegression_data4_20}
        \end{subfigure}%
        \caption{An overview of important pages. (a) MLLM-agnostic detector page and credits indicator. (b) MLLM-aware detector page. (c) Result page for image test. (d) MLLM-aware detector result. (e) Statistics page. (f) Feedback page.}\label{fig: pages}
\end{figure*}

The \textbf{SynthGuard Frontend} is a responsive single-page web application (SPA) developed with {React v18}. Its component-based design enables modular reuse, rapid prototyping, and consistent performance across devices. The interface interacts with the backend through secure RESTful APIs, ensuring seamless operations such as media uploads and prediction visualization without page reloads.

Designed for \textbf{usability and transparency}, the frontend includes modules for media upload, detector selection, feedback submission, and model comparison, updated dynamically via React state management and asynchronous APIs. It supports \textbf{low-latency inference visualization}, enabling real-time display of model outputs and confidence scores.
SynthGuard also incorporates structured feedback and credit-based access to facilitate controlled experiments and reproducible research. The interface remains extensible for future multimodal integrations and analytics modules focused on user trust and interpretability (see Fig.~\ref{fig:ui} for the Landing Page design).

\subsection{Account}
The \textit{Account} module enables secure user registration and authentication. During sign-up, users provide basic information—name, email, position, region, and a password/confirmation pair. All passwords are hashed using \texttt{bcrypt}, ensuring no plaintext credentials are stored. The registration and login pages use a glass-style card layout with stepwise validation, password-strength feedback, and a fully responsive design that works smoothly on both mobile and desktop screens
Each new account starts with \textbf{20 usage credits}, which serve as a unified metric for both image and audio evaluations. The remaining balance is shown through the \textit{Credit Tracker} on the Models page (see Fig.~\ref{fig: pages} (a)). When credits reach zero, the interface displays an automated alert prompting the user to request more. Users can email the M2 Lab from the same address linked to their SynthGuard account, briefly describing their usage and the number of credits needed. After review, the team restores or extends credits as appropriate.
From a security and governance standpoint, each account is tied to a unique \texttt{UserID}, preventing duplicates and ensuring clear attribution. Region and position are collected only at a minimal level to study adoption trends across geographic and professional groups without intrusive profiling. Required fields, strong-password rules, and secure transmission of registration data help verify legitimate users, reduce automated misuse, and support responsible, data-driven evaluation of SynthGuard’s deployment.



\subsection{MLLM-Agnostic/-Aware Detector Pages}
\textbf{MLLM-Agnostic Detectors.} SynthGuard includes dedicated pages for image and audio deepfake detection powered by a diverse suite of AI models such as \textit{Xception}~\cite{chollet2017xception}, \textit{EfficientNet-B4}~\cite{tan2019efficientnet}, \textit{ViT-B/16}~\cite{dosovitskiy2021vit}, \textit{F3Net}~\cite{qian2020thinking}, and \textit{PG-FDD}~\cite{lin2024pgfdd}. All of them are obtained from \cite{lin2025ai,ren2024improving}. These models collectively enable reliable multimodal inference by leveraging both spatial and frequency-aware features, ensuring robust detection across diverse generative architectures and manipulation styles.  
The Detector page accepts standard image formats including \texttt{.png}, \texttt{.jpg}, and \texttt{.avif}, as well as audio formats like \texttt{.wav} and \texttt{.mp3}. Users can upload or drag-and-drop files into the detection interface, view progress indicators, and obtain predictions in real time. As shown in Fig.~\ref{fig: pages} (a), the page also integrates a real-time \textbf{credit tracker} that displays available usage credits and consumption per inference, ensuring transparent access control for research users.

\textbf{MLLM-Aware Detectors}. The integrated MLLM (Multimodal Large Language Model) workspace enables reasoning-based assistance for uploaded media, leveraging Qwen-VL-Chat and Qwen-Audio to provide contextual analysis and explanations (see Fig.~\ref{fig: pages} (b) for how each detector icon was selected). 
To use the workspace, the user begins by selecting one of the available multimodal models—such as {Qwen-VL-Chat}, {InternVL-Chat-V1.5}, {LLaVA-NeXT}, {Whisper (Audio)}, or {Qwen2-VL-2B}. The interface allows users to upload an image (\eg, \texttt{.jpg}, \texttt{.png}) or an audio file (\eg,  \texttt{.wav}, \texttt{.mp3}) depending on the selected mode. After entering a query in the prompt bar (e.g., “Is this image real?” or “Does this audio sound authentic?”), the system processes the input and provides real-time feedback through progress indicators. Once inference completes, the selected model generates a detailed response, and the user may continue the conversation to perform deeper analysis, request clarifications, or run additional multimodal checks.


\subsection{Detection Results}
Once the backend completes inference, results are displayed in an organized and visually intuitive format. Each prediction card presents the uploaded image or audio metadata alongside the predicted authenticity label (\emph{real} or \emph{fake}) and confidence scores expressed as percentages. For image-based detection, the UI uses a radial confidence gauge with green for real and orange-red for fake predictions. Multi-face inputs include cropped thumbnails and annotated bounding boxes, while audio predictions include a waveform view with prediction overlays.
These visualizations help users interpret model outcomes quickly and clearly. See Fig.~\ref{fig: pages} (c) for the image result page. 
In addition to the primary deepfake detectors, the platform integrates outputs from the multimodal LLM-aware module (MLLM), which provides semantic descriptions rather than authenticity judgments. Unlike the core detectors, the MLLM does \emph{not} classify an input as real or fake. Instead, it produces a detailed natural-language interpretation of visual or auditory content. For image inputs, the MLLM identifies high-level attributes such as the subject’s appearance, clothing colors, facial features, visible objects, and scene context. When multiple people are present, the MLLM generates per-face attribute summaries to help users understand what the system perceives in each region of the image. For audio inputs, the model can describe speaker characteristics, tone, and contextual cues without issuing authenticity labels. These semantic explanations complement the detector predictions by offering human-readable insights into the content itself, thereby enhancing transparency and contextual understanding of the system’s outputs (see Fig.~\ref{fig: pages} (d)).

\FloatBarrier   




\subsection{Statistics}
SynthGuard features a React-built interactive {dashboard} that visualizes platform usage and research metrics. Charts summarize total users, number of predictions made, and the distribution of real versus fake detections. Additional widgets display model utilization rates, per-category activity, and geographical distribution of users. The dashboard is implemented using lightweight, high-performance chart libraries and dynamically updates based on MySQL aggregates returned by the backend analytics API (see Fig.~\ref{fig: pages} (e) for an example of the statistics interface).


\subsection{Feedback Page}
SynthGuard incorporates a dedicated \textit{Feedback Page} to collect structured user insights on usability, model performance, and system improvements. The form captures both qualitative and quantitative data to guide future platform refinements.

Users specify which \textit{models} (\texttt{Xception}, \texttt{EfficientNet-B4}, \texttt{ViT-B/16}, \texttt{F3Net}, \texttt{CORE}, \texttt{UCF}, \texttt{UnivFD}, etc.) and \textit{file formats} (image, audio, or video) they used, followed by the model they perceived as most \textit{accurate}. Additional sections gather feedback on useful \textit{features}, desired \textit{improvements}, and an overall \textit{experience rating}. Open-ended fields invite creative suggestions such as mobile app extensions or video-level detection. Demographic inputs, including \textit{user role} and prior exposure to deepfake tools, provide context for response analysis.

All submissions are securely transmitted to the backend for anonymized aggregation. This data helps the Purdue M2 Lab assess usability trends, refine detection workflows, and prioritize updates. Refer to Fig.~\ref{fig: pages} (f) for an illustration of the Feedback Page interface.




\section{SynthGuard Backend}
\begin{figure*}[t]
  \centering
  \includegraphics[width=\textwidth]{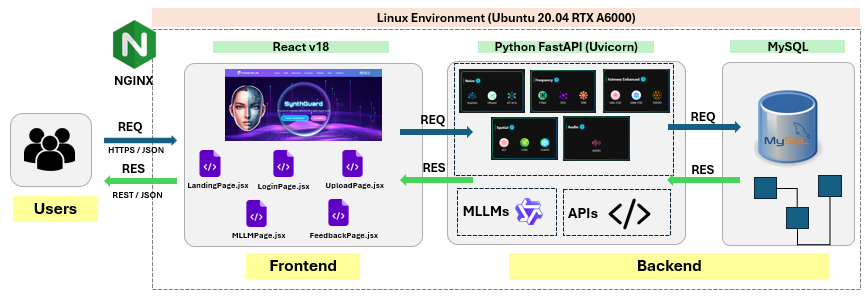}
  \caption{SynthGuard platform architecture. The system operates end-to-end across a full-stack pipeline: 
(1) Users interact with the React v18 frontend over HTTPS/JSON to upload images or audio and manage accounts; 
(2) Requests are routed through an NGINX reverse proxy to the FastAPI backend hosted on a Linux server (Ubuntu 20.04 with NVIDIA RTX A6000); 
(3) The backend invokes integrated multimodal LLM-aware detectors and model APIs to analyze submitted media; 
(4) Results, predictions, and metadata are returned to the frontend; 
(5) User information, credit balances, and inference logs are persisted in a MySQL database.}

  \label{fig:upload}
\end{figure*}
The \textbf{SynthGuard Backend} powers the inference engine, authentication, and analytics layers of the platform. Built with {Python FastAPI}, it leverages asynchronous I/O for high-throughput performance and integrates with PyTorch-based detectors and multimodal LLM modules for real-time inference.
As shown in Fig.~\ref{fig:upload}, the system follows a three-tier architecture—frontend, backend, and database deployed within Purdue’s on-premise Linux environment. {Uvicorn} handles concurrent execution efficiently, while {NGINX} provides secure reverse proxying and request routing. This modular design allows new APIs, inference models, and visualization features to be added independently, supporting both scalability and maintainability as user concurrency and data volume grow.

\subsection{Backend Technology Stack}
The backend leverages {FastAPI} for RESTful API routing and {Uvicorn} as the ASGI server for concurrency and efficiency. Deep learning detectors are implemented using {PyTorch}, with pre-trained model weights stored as \texttt{.pth} files. Other critical components include {SQLAlchemy} for database operations, {Pydantic} for schema validation, and {JWT} for secure user authentication. The backend architecture is stateless, allowing scaling across multiple worker nodes under load.

\subsection{Detectors}
\textbf{SynthGuard} integrates multiple categories of deepfake detectors and multimodal models to ensure robust, cross-modal analysis across image and audio inputs. Note that all image detectors are trained and evaluated on the \textit{AI-Face} benchmark \cite{lin2025ai} to maintain consistency and comparability across domains.  

\subsubsection{MLLM-Agnostic Image Detectors}
\begin{itemize}
    \item \textbf{Backbone-only Detectors:} Methods that rely solely on CNN or Transformer backbones to learn discriminative spatial features, without frequency priors or handcrafted artifact modeling. They provide strong architectural baselines for RGB-based forgery detection.
    \begin{itemize}
        \item \textbf{Xception}~\cite{chollet2017xception}: A deep convolutional neural network that employs depthwise separable convolutions to reduce parameters and computation while maintaining high accuracy.  
        \item \textbf{EfficientNet-B4}~\cite{tan2019efficientnet}: Introduces compound scaling to balance network depth, width, and resolution, achieving high efficiency across detection tasks. 
        \item \textbf{ViT-B/16}~\cite{dosovitskiy2021vit}: A Vision Transformer model that processes $16\times16$ patches with a transformer encoder, enabling strong global feature learning for forgery detection. 
    \end{itemize}

    \item \textbf{Frequency-Based:} Detectors that analyze spectral and phase-domain cues to reveal high-frequency artifacts from generative pipelines, including periodic irregularities and upsampling distortions.
    \begin{itemize}
        \item \textbf{F3Net}~\cite{qian2020thinking}: Integrates Frequency-aware Decomposition and Localized Frequency Statistics to capture subtle spectral artifacts.
        \item \textbf{SPSL}~\cite{liu2021spsl}: Leverages spatial–phase spectrum features to detect up-sampling artifacts and improve generalization.
        \item \textbf{SRM}~\cite{luo2021srm}: Extracts high-frequency residual noise from RGB and frequency domains to enhance robustness.  
    \end{itemize}

    \item \textbf{Spatial-Based:} Models that focus on pixel-level, structural, and texture inconsistencies in forged images, emphasizing spatial-domain cues and representation regularization.
    \begin{itemize}
        \item \textbf{UCF}~\cite{yan2023ucf}: Uses multi-task disentanglement to identify shared forgery patterns across generative models.
        \item \textbf{CORE}~\cite{ni2022core}: Enforces consistency across augmented representations through cosine regularization.
        \item \textbf{UnivFD}~\cite{ojha2023univfd}: A universal fake detector using a frozen CLIP ViT-L/14 feature extractor with a fine-tuned linear classifier.
    \end{itemize}

    \item \textbf{Fairness-Enhanced:} Detectors augmented with fairness-driven optimization to reduce demographic bias and improve equitable performance across groups.
    \begin{itemize}
        \item \textbf{DAW-FDD}~\cite{ju2024fairness}: Uses CVaR loss to reduce demographic bias and improve fairness. 
        \item \textbf{DAG-FDD}~\cite{ju2024fairness}: A distributionally robust variant that enhances fairness without demographic annotations. 
        \item \textbf{PG-FDD}~\cite{lin2024pgfdd}: Integrates disentanglement and fairness-driven optimization to preserve generalization across domains.  
    \end{itemize}
\end{itemize}

\subsubsection{MLLM-Agnostic Audio Detectors} 
A lightweight \textit{CNN-based} architecture optimized for synthetic speech detection~\cite{ren2024improving}. The model analyzes mel-spectrograms, harmonic–noise structures, and phase coherence to capture spectral distortions characteristic of voice cloning and vocoder-based generation, enabling robust cross-speaker and cross-accent detection.

\subsubsection{MLLM-Aware Detectors}
\textit{Qwen-VL-Chat}~\cite{qwen-vl-chat}, 
\textit{LLaVA-NEXT-13B}~\cite{llava-next}, 
and \textit{InternVL-Chat-V1.5}~\cite{internvl-chat}
are used to perform cross-modal reasoning and semantic verification across text and images, consistent with the ForensicsBench evaluation protocol~\cite{forensicsbench2025}.
For audio analysis, the system adopts a hybrid pipeline in which the \textit{Whisper} model~\cite{whisper} provides robust speech transcription, and the resulting text is analyzed using \textit{Qwen2-VL-2B}~\cite{wang2024qwen2vlenhancingvisionlanguagemodels} to assess semantic consistency and detect potential audio forgeries. This design enables reliable audio reasoning without requiring a dedicated audio–language MLLM.

\smallskip
\textbf{Modular Integration.}
SynthGuard is designed with a modular plugin-style detector interface. New image, audio, or multimodal detector, whether backbone-driven, frequency-aware, fairness-optimized, or MLLM-based, can be integrated with minimal code changes. This ensures long-term extensibility as new generative models and forensic techniques continue to emerge.


\subsection{Codebase Structure}
The \textbf{SynthGuard} platform is divided into independent {frontend} and {backend} components. The backend, built with {Python FastAPI}, includes modular directories for APIs, models, utilities, and database operations, while the frontend (React) manages visualization and interaction. Development uses {VS Code}, {MySQL Workbench}, and {Postman} for coding, database design, and API testing, with unit tests ensuring reliability. All APIs, SQL scripts, configurations, and functions are fully documented and version-controlled for reproducibility.

\subsection{Hosting}
The system is hosted on a {Linux server (Ubuntu~20.04~LTS)} equipped with high-performance NVIDIA GPUs for deep learning inference. The deployed configuration utilizes \textbf{eight NVIDIA RTX~A6000} GPUs, each based on the Ampere architecture and featuring \textbf{10,752~CUDA cores} and \textbf{336~Tensor cores}. The compute node is powered by {dual Intel Xeon~Gold~6226R} processors (2.90\,GHz, 16~cores per socket, 64~threads total), providing balanced throughput for data preprocessing and multi-model execution. The backend is deployed as systemd-managed {Uvicorn} worker processes behind an {NGINX} reverse proxy, ensuring load balancing, TLS security, and continuous uptime within Purdue’s IT infrastructure.

\subsection{Database}
The {Database} is implemented using a structured {MySQL} database that manages all persistent information within the SynthGuard platform. Core tables include \texttt{USERS}, \texttt{UPLOADS}, \texttt{PREDICTIONS}, \texttt{FEEDBACK}, \texttt{CREDITS}, and \texttt{MODEL\_LOGS}, each maintaining detailed records of system interactions and inference results. Every entry is associated with timestamps, model identifiers, and modality types (image or audio), enabling reproducible research and statistical evaluation.

SynthGuard upholds strong principles of {data governance} and transparency. User permission is explicitly requested before collecting any image or audio data, and all submissions are securely transmitted and anonymized. The aggregated data assists the Purdue M2 Lab’s AI research team in refining and benchmarking evolving deepfake detection models. The results derived from this layer directly inform the analytical dashboards and statistical visualizations presented to the user community.


\section{Conclusion}
The \textbf{SynthGuard} platform integrates robust frontend design, scalable backend computation, and a secure, analytics-driven database to provide a unified environment for deepfake detection and multimodal forensic research. Its modular and extensible architecture ensures reproducibility, transparency, and cross-domain adaptability.

Beyond architectural robustness, {SynthGuard} offers several unique advantages. The platform unifies image-, audio-, and text–image semantic verification within a single pipeline, enabling both traditional forgery detection and higher-level MLLM-aware reasoning. 
SynthGuard further incorporates fairness-enhanced detectors, credit-managed experiment workflows, and interpretable probability calibration, allowing researchers to rigorously evaluate performance across demographic and generative domains.

The system’s plugin-style detector interface enables rapid integration of new detectors, including backbone-only, frequency-based, spatial-based, and multimodal LLM models. This design ensures long-term extensibility as generative techniques continue to evolve. Taken together, SynthGuard provides a novel, unified, and research-grade framework that advances the state of deepfake forensics and supports reproducible, ethically grounded multimodal analysis.

\subsection{Limitations}
While {SynthGuard} provides a unified and extensible framework for multimodal deepfake forensics, several limitations remain. First, the MLLM-aware detectors integrated into the system are not specifically trained for deepfake detection; rather, they are repurposed from general multimodal reasoning tasks. As a result, their semantic inconsistency analysis may not fully capture the fine-grained visual or temporal artifacts characteristic of modern generative pipelines.

Second, the MLLM-agnostic image detectors are primarily trained on face-centric datasets. Although this provides strong performance in facial forgery detection, these models do not generalize to broader categories of AI-generated images (e.g., scenes, objects, artwork). Their applicability is therefore limited outside facial domains.

Finally, the current platform does not incorporate any video-based deepfake detectors. All detectors operate on static frames or single images, and SynthGuard does not yet model temporal cues such as motion inconsistency, frame-level transitions, or temporal frequency artifacts that are critical for video-level forensics. 

\subsection{Future Work}
Future development of {SynthGuard} will focus on broadening both the modality coverage and the detector diversity. A key direction is the integration of detectors capable of identifying general AI-generated media beyond facial forgeries, as well as incorporating deepfake-specific MLLM-aware models that offer stronger semantic and context-level reasoning. Expanding the platform to include video- and text-based forgery detectors is also a priority, enabling temporal analysis, cross-modal alignment, and unified multimodal forensics.

In parallel, we envision exploring a transition toward cloud-native infrastructure using AWS; however, this remains an aspirational direction as resource provisioning costs must be carefully evaluated. As platform usage and user concurrency grow, the team plans to conduct small proof-of-concept deployments to assess the feasibility of leveraging services such as AWS S3 for scalable media storage, AWS Aurora for high-availability database management, and containerized workflows for improved concurrency. These exploratory efforts will help determine the practicality and long-term value of adopting cloud-native components.

\section*{Acknowledgements}
This work is supported by the U.S. National Science Foundation (NSF) under grants IIS-2434967 and TI-2448500, and the National Artificial Intelligence Research Resource (NAIRR) Pilot and TACC Lonestar6. The views, opinions and/or findings expressed are those of the author and should not be interpreted as representing the official views or policies of NSF and NAIRR Pilot.

\bibliographystyle{IEEEtran}
\bibliography{main}

\end{document}